\documentclass[superscriptaddress,showpacs,twocolumn,prl,floatfix]{revtex4-1}
\usepackage{bm}
\usepackage{graphicx}

\def\be{\begin{equation}}
\def\ee{\end{equation}}
\def\ber{\begin{eqnarray}}
\def\eer{\end{eqnarray}}

\def\nn{\nonumber}

\usepackage [english]{babel}
\usepackage{color}

\begin{document}
\title{Spin current generation from Coulomb-Rashba interaction in semiconductor bilayers}
\author{M. M. Glazov}
\author{M. A. Semina}
\affiliation{Ioffe Physical-Technical Institute of the Russian Academy of Sciences, 
St. Petersburg, 194021 Russia}
\author{S. M. Badalyan}
\affiliation{School of Physics, Astronomy, and Computational Sciences, George Mason University, Fairfax Virginia 22030, USA}
\affiliation{Department of Physics, University of Antwerp, Groenenborgerlaan 171, B-2020 Antwerpen, Belgium}
\affiliation{Department of Physics, University of Regensburg, 93040 Regensburg, Germany}
\author{G. Vignale}
\affiliation{Department of Physics and Astronomy, University of Missouri, Columbia, Missouri 65211, USA}
\begin{abstract}
{Electrons in double-layer semiconductor heterostructures
    experience a special type of spin-orbit interaction which arises
    in each layer  from the perpendicular  component of the Coulomb
    electric field created by electron density fluctuations in the
    {\it other} layer.  We show  that this interaction, acting in
    combination with the usual  spin-orbit interaction,  can generate
    a spin current in one layer when a charge current is driven in the
    other.  This effect is symmetry-wise distinct from the spin Hall
    drag.  The spin current is not, in general, perpendicular to the
    drive current.}
\end{abstract} 
\maketitle

Spin-orbit interactions in semiconductors  are traditionally studied
within a one-electron picture,  but there are instances in which
electron-electron interactions latch onto spin-orbit interactions to
produce intriguing effects, which may lead to the creation of
radically new spin-based electronic
devices{~\cite{as,spin_book,fmesz}.}   Of particular interest is the
generation of spin currents and spin accumulation by an electric
current, through the so-called {\it spin Hall
  effect}~{\cite{SHE_gen}}.  Recently, a special type of spin Hall
effect was predicted to occur in double-layer heterostructures, i.e.,
two parallel quantum wells separated by an essentially impenetrable
potential barrier,  with a quasi-two dimensional electron gas in each
layer.   This effect, called ``spin Hall drag"{~\cite{badalyan-2009}},
consists of the generation of transverse spin accumulation in one
layer by an electric current flowing along the other layer and is
caused by the component of the Coulomb electric field parallel to the
two layers. In this Letter we study a different  and novel effect,
which is driven by the component of the Coulomb electric field 
{\it perpendicular} to the layers.  This perpendicular field creates
an inhomogeneous Rashba spin-orbit interaction, with spatial variation
in the plane of the layer~\cite{footnote}. We refer to this
interaction as the  ``Coulomb-Rashba interaction". 

\begin{figure}[hptb]
\includegraphics[width=0.7\linewidth]{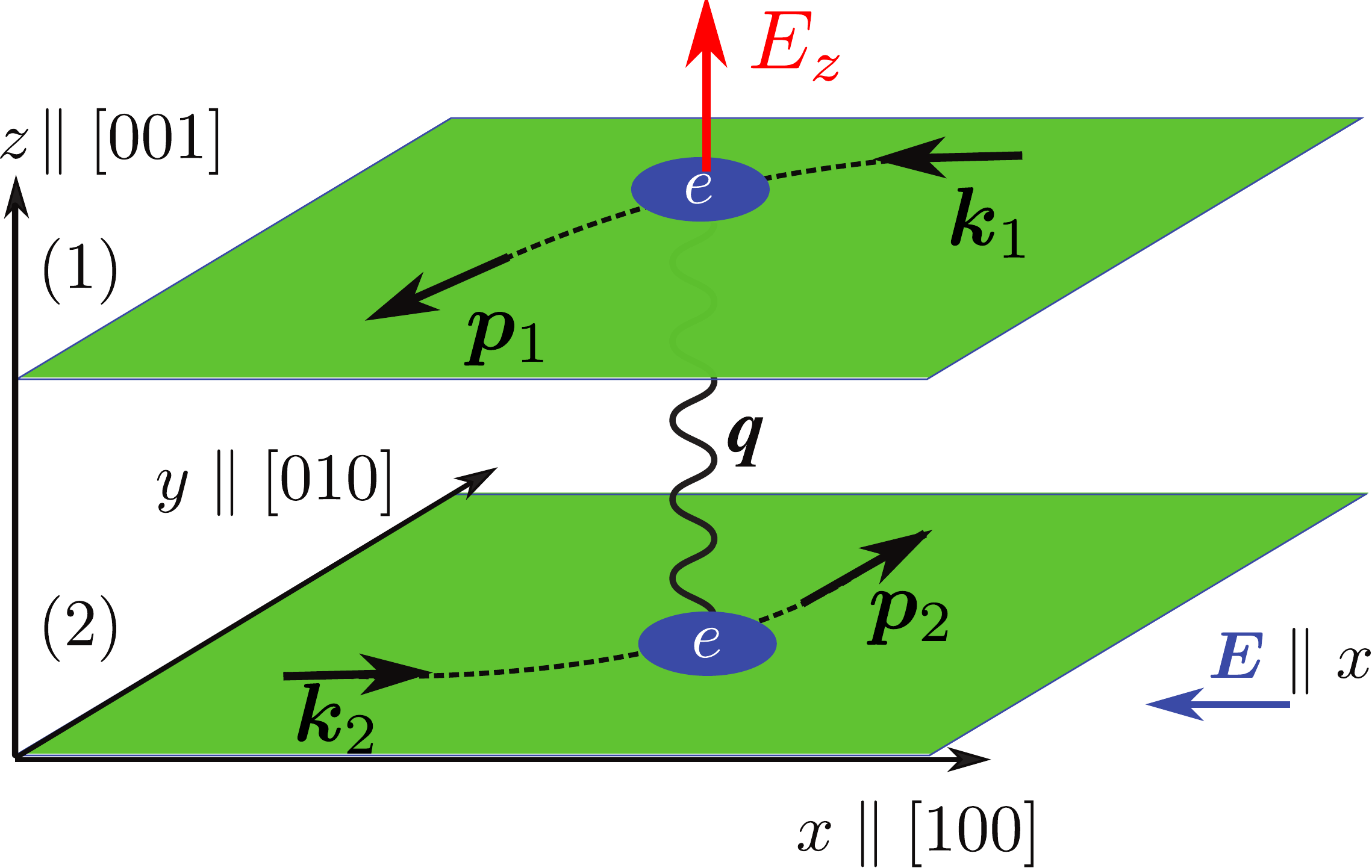}
\caption{Scheme of the device under study: an electric field, $\bm E$,
  is applied to the active layer (2), a spin current is generated in
  the passive layer (1). Blue circles depict electrons, and the wavy
  line shows the inter-layer Coulomb interaction. Arrows marked $\bm
  k_1$, $\bm k_2$ and $\bm p_1$, $\bm p_2$ are the wave vectors in the
  initial and final states. The vertical arrow emphasizes the relevant
  component of the interlayer Coulomb field.} 
\label{fig:device}
\end{figure}

The system under study is shown in Fig.{~\ref{fig:device}}.  A steady
electric current is driven in the active layer 2. We show that the
interplay of the {Coulomb-Rashba interaction with the ordinary
  cubic Dresselhaus spin-orbit interaction} (characteristic of
semiconductors of the zincblende structure),  provides a new mechanism
for the generation of a spin current in the passive layer 1.
This is particularly remarkable in view of the fact that
  ordinary spin Hall effect and spin Hall drag are {\em suppressed} by
  spin precession. But, in this case the presence of spin precession
  is absolutely essential to the effect.

\begin{figure}
\includegraphics[width=\linewidth]{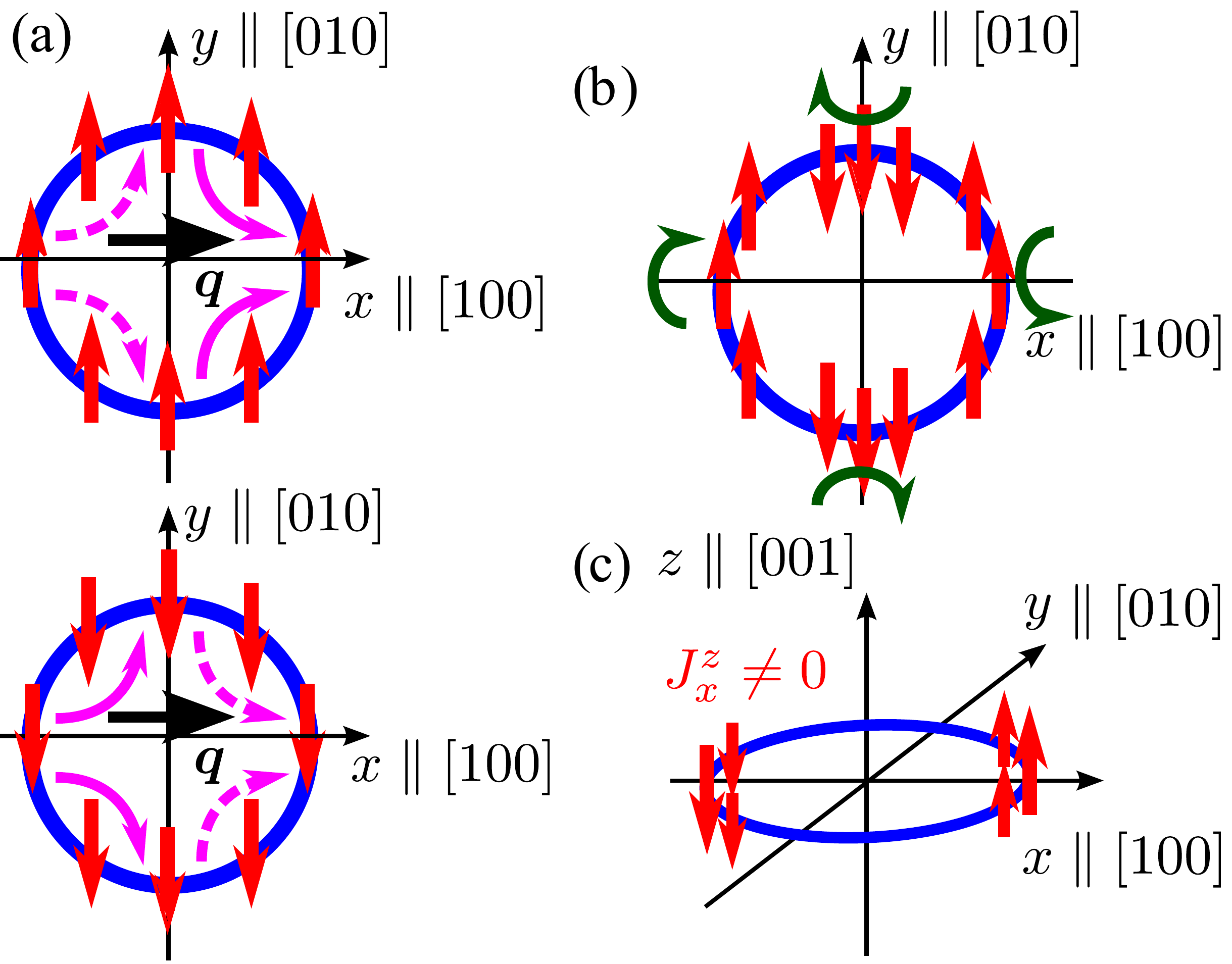}
\caption{Schematic illustration of the spin current generation in the passive layer~\cite{note:tar}. The circle is the Fermi surface, the arrows are the electron spins. Panel (a) shows the scattering stage of the process: top part shows electrons with spin parallel to the $y$ axis, bottom part shows electrons with spin antiparallel to the $y$ axis. The arrows show the spin-dependent scattering process: solid arrows indicate the stronger transitions, having $k_x+p_x>0$ for $s_y>0$ and $k_x+p_x<0$ for $s_y<0$,  while dashed arrows indicate the weaker transitions. Scattering processes that increase the $x$ component of the wave vector (i.e. with $q_x>0$) dominate due to the current flowing in the active layer. (b) Precession stage of the process. The resulting spin distribution after the scattering contains second angular harmonics. Green arrows demonstrate the spin precession caused by the Dresselhaus field $\bm \Omega_{\bm k}$, Eq.~(\ref{Dress}). Electron spins with opposite wave vectors precess in opposite directions. The resulting dipolar distribution of electron spins is presented in the panel (c).}
\label{fig:model}
\end{figure}

The spin current generation can be understood as a two-stage process, which is schematically described in Fig.~\ref{fig:model}~{\cite{note:tar}}. 
In the first stage, the steady current of electrons  in layer 2  induces a quadrupolar spin distribution in layer 1.  To understand qualitatively how this comes about, observe that Coulomb collisions between electrons in the two layers take place on average with a positive momentum transfer from layer 2 to layer 1 along the $x$ axis. 
Moreover, due to the spin-dependent terms in the scattering rate (see Eq.~\ref{Coulomb1} below), the scattering efficiency depends on the relative orientations of electron spin and its initial and final wavevectors, denoted by $\bm k$ and $\bm p$ respectively.  Consider electrons with spins parallel or antiparallel to the $y$ axis: $s_y>0$ and $s_y<0$ respectively.   For $s_y>0$  the scattering rate is maximal for $k_x+p_x>0$ and minimal for $k_x+p_x<0$.   For $s_y<0$  the situation is reversed: the scattering rate is maximal for $k_x+p_x<0$  and minimal for  $k_x+p_x>0$.  The strongest transitions are marked by solid arrows in Fig.~\ref{fig:model}(a).   As a result, states with large values of $|k_y|$  ($|k_y|\sim k_F$)  become depleted of $s_y>0$ electrons and filled with $s_y<0$ electrons.  At the same time, states with large values of $|k_x|$  ($|k_x|\sim k_F$)   are depleted of  $s_y<0$ electrons  and filled with $s_y>0$ electrons.
Hence, a quadrupolar spin distribution is formed, as shown in Fig.~\ref{fig:model}(b). 

The second stage of the spin current generation is related to the spin
precession in the cubic Dresselhaus field.  This is also illustrated
in Fig.~\ref{fig:model}(b), where the green arrows show the direction
of the spin precession induced by the cubic Dresselhaus field, $\bm
\Omega_{\bm k}$, (see Eq.~\ref{Dress} below).  The field tilts the
spins out of the $x-y$ plane, thus creating a dipolar distribution of
the $z$-component of the spin as shown in  Fig.~\ref{fig:model}(c).
As a result, a $z$-spin current is formed.  The characteristic
$C_{2v}$ symmetry of systems with both Rashba and Dresselhaus
interactions causes the spin current to be parallel to the driving
electric field $\bm E$, when $\bm E$ is along one of the cubic $[100]$
or $[010]$ axes.  {Notice that this is completely different from the
  spin Hall drag current, which is always perpendicular to the
  electric field.  However, the spin current can also be made
  perpendicular to $\bm E$, by orienting $\bm E$ is along one of the
  principal axes $[110]$ or $[\bar110]$.}

In the rest of the paper we present some details of our theoretical analysis, and provide a numerical estimate for the size of the effect.  The salient conclusions are as follows: (i)  The spin transresistivity is proportional to $(T/E_F)^2$ which is characteristic of Coulomb drag phenomena~\cite{gramila,jauho,badalyan-2009} (ii) It is inversely proportional to the fourth power of the interlayer separation and  (iii) It is parametrically stronger than the spin Hall drag~\cite{badalyan-2009} in the clean limit since the transresistivity is proportional to $\Omega \tau$, whereas the $T^2$ contribution to the spin-Hall drag is independent of $\tau$.

\noindent{\it Theory --}  The part of the ``spin-orbit dressed" Coulomb interaction which is relevant to the effect described in this paper has matrix elements~{\cite{glazov2009,glazov2009a,sherman_rev}}
\begin{eqnarray}\label{Coulomb1}
&&M(\bm k_1s_1, \bm k_2s_2 \to \bm p_1s_1', \bm p_2s_2') =  \{V_q-\frac{\lambda^2}{2}V_q (q+q_s) \nn\\
&&\langle \chi_{s_1'}\chi_{s_2'} |[\hat{\bm \sigma}_1 \times (\bm p_1+\bm k_1)]_z  -  [\hat{\bm \sigma}_2 \times (\bm p_2+\bm k_2)]_z |\chi_{s_1}\chi_{s_2}\rangle\} \nonumber\\
&&\delta_{\bm k_1+\bm k_2,\bm p_1 + \bm p_2}\,,
 \end{eqnarray}
where the subscripts $1$ and $2$ refer to electrons in layers $1$ and $2$, respectively, $\hat{\bm \sigma}_1$ and $\hat{\bm \sigma}_2$ are the spin operators of the first and second electron acting on spinors $|\chi_{s_1}\rangle$,$|\chi_{s_1'}\rangle$ and  $|\chi_{s_2}\rangle$,$|\chi_{s_2'}\rangle$, respectively. The normalization area is set to unity.  Here  $\bm q = \bm p - \bm k$ is the transferred wave vector,
\[
V_q=\frac{2\pi e^2}{ \kappa } \frac{q\mathrm e^{-qL}}{(q+q_s)^2-q_s^2\mathrm e^{-2qL}}
\]
is the Fourier transform of the screened Coulomb interaction between the layers, $L$ is the distance between the layers, $\kappa$ is a background dielectric constant and $q_s$ is the screening wave vector.  The strength of the spin-orbit interaction is controlled by the ``effective Compton wavelength" $\lambda$ for the semiconductor:
\begin{equation}
\label{xi}
\lambda^2 = -\frac{P^2}{3}\frac{\Delta(2E_g+\Delta)}{E_g^2(E_g+\Delta)^2}\,,
\end{equation}
where $P$ is the  Kane parameter~\cite{Winkler} and $E_g$ and $\Delta$ are the band-gap and the spin-orbit splitting of the valence band.  In GaAs $\lambda^2 \approx 5$ \AA$^2$. 
Notice that, in writing Eq.~(\ref{Coulomb1})  we have taken into account only the Rashba-like contribution arising from the component of the Coulomb field perpendicular to the layers.  The additional term  coming from the component of the Coulomb field parallel to the layers was discussed in Refs.~\onlinecite{glazov2010,boguslawski, badalyan-2009} and is not shown in Eq.~(\ref{Coulomb1}).

As discussed in the introduction and in the caption of Fig.{~\ref{fig:model}}, a steady current driven in the active layer (2) produces, via the ``spin-orbit dressed" Coulomb interaction, a quadrupolar distribution of spin in the passive layer.  The spin generation rate in the passive layer can be calculated by means of  the Fermi golden rule with the matrix element of interlayer electron-electron interaction given by Eq.~(\ref{Coulomb1})~\cite{damico,glazov04a}.  To first order in $\lambda^2$ this gives
\begin{eqnarray}\label{Qs}
&&\bm g_{\bm k} = 
\frac{4 \pi\hbar e\tau}{m k_B T} \sum_{{\bm k}'
{\bm p} {\bm p}'}\delta_{{\bm k} + {\bm k}',\: {\bm p} + {\bm
p}'} \delta(\mathcal E_k + \mathcal E_{k'} - \mathcal E_p - \mathcal E_{p'}) \times 
\nonumber\\
&&U_{\bm k - \bm p}[(\bm p+\bm k)\times \bm {\hat z}]  (\bm E \cdot (\bm p - \bm k))f_{k}f_{k'}
(1-f_{p})(1-f_{p'})\,,
\end{eqnarray}
where $U_q \equiv \lambda^2 (q+q_s)|V_q|^2$,  $\bm E$ is the electric field acting on the carriers in the active layer, $m$ is the effective electron mass, $k_B T$  is the temperature measured in the energy units, $\mathcal E_k = \hbar^2k^2/(2m)$ is the electron dispersion, $f_k$ is the Fermi-Dirac distribution function and $\bm {\hat z} $ is the unit vector normal to the layers.  We have also assumed that equilibrium densities, effective masses and Fermi energies are the same in the two layers.   
In deriving Eq.~(\ref{Qs}) we took into account only linear-in-$\bm E$ nonequilibrium correction to the distribution function.  It is easy to show that, due to the presence of the two factors $ \bm E \cdot (\bm p - \bm k)$ and $(\bm p+\bm k)\times \bm {\hat z}$, the angular dependence of $g_{\bm k}$ is an angular harmonic of order 2, i.e. we have
$g_{x,{\bm k}}\propto \sin 2{{\varphi_{\bm k}}}$ and $g_{y,{\bm k}}\propto -\cos 2{{\varphi_{\bm k}}}$, where $\varphi_{\bm k}$ is the angle of $\bm k$ with the $x$ axis: this is the quadrupolar pattern of spin generation mentioned in the introduction and described in Fig.~\ref{fig:model}.

The spin dynamics in the passive layer is governed by the spin-orbit
splitting of the energy spectrum. Since the first harmonic components
of the spin splitting (arising from linear-in-$\bm k$-terms) do not
result in a $dc$ spin Hall
current~\cite{chalaev:245318,dimitrova:245327,tarasenko06e}, we only
take into account the third angular harmonics of the $k^3$ Dresselhaus
term which is inevitable in any zincblende structure. 
In cubic axes with $x\parallel [100]$ and $y\parallel [010]$ these have the form 
\begin{equation}
\label{Dress}
\Omega_{x,\bm k} = -\Omega_3\cos{3\varphi_{\bm k}}, \quad \Omega_{y,\bm k} = -\Omega_3\sin{3\varphi_{\bm k}},
\end{equation}
where $\Omega_3=\gamma_c k^3/(2\hbar)$, $\gamma_c$ is the bulk Dresselhaus constant. 
The electron spin distribution function, $\bm s_{\bm k}$, is determined by
a kinetic equation, which in the steady state takes the form~\cite{tarasenko06e}
\begin{equation}\label{kin:s}
\bm s_{\bm k} \times \bm \Omega_{\bm k} + \frac{\bm s_{\bm k}}{\tau} = \bm g_{\bm k}\,.
\end{equation}
Here we have assumed that all spin-independent scattering processes can be characterized by a single relaxation time $\tau$. 
The solution of Eq.~(\ref{kin:s}) is
\be\label{SZK}
s_{z,{\bm k}}=\frac{(\bm \Omega_{\bm k} \times {\bm g}_{\bm k})_z}{\Omega_{\bm k}^2+1/\tau^2}\,.
\ee
We note that such a simple form of the solution results from keeping only the third angular harmonics in $\Omega_{\bm k}$. The solution in the general case can be constructed following Refs.~\onlinecite{tarasenko06e,PhysRevB.77.165341}: in such a case the overall spin will be smaller due to faster spin relaxation caused by the linear-in-$\bm k$ Dresselhaus term, and the spin distribution may have more complicated form due to the anisotropy of the spin splitting.

By definition, the current density of spin $z$ component is
\begin{equation}
\label{Jspin}
\bm J^z= \sum_{\bm k} 
s_{z,\bm k} \bm  v_{\bm k}\,,
\end{equation}
where $\bm  v_{\bm k} = \hbar \bm k/m$ is the electron velocity. 
Substituting the expression for ${\bm g}_{\bm k}$, obtained from Eq.~(\ref{Qs}),  into Eq.~(\ref{SZK}) for $s_{z,{\bm k}}$, and then $s_{z,{\bm k}}$ in Eq.~(\ref{Jspin}), we arrive at our main result
\begin{eqnarray}\label{SpinCurrentDensity}
\bm J^z &=&
 -\frac{2\pi\hbar e\tau^3}{m k_B T} \sum_{{\bm k}{\bm k}'
{\bm p} {\bm p}'} \bm v_{\bm k} \ \delta_{{\bm k} + {\bm k}',\: {\bm p} + {\bm
p}'} \delta(\mathcal E_k + \mathcal E_{k'} - \mathcal E_p - \mathcal E_{p'}) 
\nonumber\\&&2U_{\bm k - \bm p}[\bm \Omega_{\bm k} \cdot (\bm p+\bm k)]  [\bm E \cdot (\bm p - \bm k)]\nonumber \\
&&{f_{k}f_{k'}
(1-f_{p})(1-f_{p'})}.
\label{Jz}
\end{eqnarray}
Equation~(\ref{Jz}) can be recast in the standard form:
\begin{equation}
\label{Jzgen}
J^z_i = G_{ij} E_j,
\end{equation}
where $G_{ij}$ is the spin drag conductivity. Our system is characterized by the  $C_{2\rm v}$ point symmetry group because we take into account (i) the Dresselhaus field, and (ii) the Rashba-like interaction associated with the perpendicular-to-plane component of the Coulomb field. In the  basis of the principal axes $x_1\parallel [1\bar 10]$ and $y_1\parallel [110]$ Eq.~(\ref{Jzgen}) can be written in terms of two independent constants $G$ and $G_1$ as
\begin{equation}
\label{calc}
J_{x_1}^z = (G+G_1) E_{y_1}, \quad J_{y_1}^z = (G-G_1) E_{x_1}.
\end{equation}
Equation~(\ref{calc}) gives the full phenomenological picture of the spin drag in the presence of the spin-orbit interaction. The drag mechanism described here produces $G\ne 0$ and $G_1=0$.  This rather peculiar situation implies that the spin current flows parallel to the electric field if the latter is applied along the $[100]$ or $[010]$ directions, but, in general,  it has both a parallel and a perpendicular component.  It is only when the electric field is along one of the principal axes that we get a pure transverse current, but even in this case it is in a sharp contrast with the  spin Hall drag, namely an electric field along $x_1$ produces a spin current along $y_1$, but an electric field along $y_1$ produces a spin current along $x_1$, rather than $-x_1$.  These unique symmetry-related features will help distinguishing the predicted new effect.


We now come to the quantitative {evaluation} of $G$. The expression~(\ref{SpinCurrentDensity}) for the spin current density is similar to the expression encountered in the calculation of the ordinary drag current~\cite{jauho}.  The main difference is that the integrand depends not only on the momentum transfer {$\bm q = \bm p - \bm k$}, but also on the sum of the initial and final momenta, {$\bm p +\bm k$}.  For simplicity we focus here on the case of
well separated layers, $k_F L \gg 1$, where the momentum transfer is typically small  ($q\sim L^{-1}\ll k_F$), and therefore one can replace {$\bm k +\bm p \approx 2 \bm k_{F}$}, making an error of order $q/k_F$. 
With this approximation, the standard method of evaluation can be
applied, and, after converting the wave vector sums into integrals,
and for low enough temperatures $k_B T \ll E_F$, we obtain 
\begin{equation}\label{G:int}
G=-\frac{\hbar^{3} e \Omega_3 k_F\tau^3}{2\pi m^2 k_B T}\int \frac{\mathrm d\bm q}{(2\pi)^2} \int_{-\infty}^{+\infty} \mathrm d\omega \frac{{\mathrm I\mathrm m}[\chi_0(q,\omega)]^2}{\sinh^2\left(\frac{\hbar\omega}{2k_B T}\right)} U_q q^2\,,
\end{equation}
where  $\chi_0(q,\omega)$ is the non-interacting density-density response function of the two-dimensional electron gas \cite{vignale:book}.
Integrating over frequency we get
\begin{equation}
\label{G}
G =-\frac{\lambda^2\Omega_3 e^5 k_F^4 \tau^3}{{4}\pi^2 \kappa^2 {\hbar^2}} \left(\frac{k_BT}{E_F}\right)^2 I_2(q_sL)
\end{equation}
with 
\begin{equation}\label{I2}
I_2(x)={\frac{2\pi^2}{3 k_FL}}\int_0^{\infty} \frac{y^3(y+x)e^{-2y}}
{[(x+y)^2-x^2e^{-2y}]^2} \mathrm dy\,,
\end{equation}
for $k_FL \gg 1$.  For $q_sL \gg 1$ the integral Eq.~(\ref{I2}) is evaluated to be $I_2(q_sL)  \approx 2.96/(k_Fq_s^3L^4)$, as a result $G \propto 1/L^4$ as in the ordinary drag effect~\cite{gramila,jauho}.

\noindent\textit{Results and discussion -- }It is instructive to estimate the drag resistivity which controls the observable spin accumulation.  This is determined as follows. First, we express the external field $E$ via the current density generated by this field in the active layer, $j=\sigma E$, where $\sigma= ne^2 \tau/m$ is the Drude conductivity of the electrons and $n$ is their density.  Then, we observe that the spin current is associated with some effective electric ``spin'' field $E_s$
\[
{\frac{2e}{\hbar}}J^z = \sigma E_s,
\]
 As a result we obtain
\[
E_s = {\frac{2eJ^z}{\hbar \sigma} = \frac{2eG}{\hbar \sigma^2}j = \rho_s j}.
\]
Hence, the drag resistivity is given by
\begin{equation}
\label{rhos}
\rho_s  = -\frac{2\hbar}{e^2}(\lambda k_F)^2 \Omega_3\tau \left(\frac{k_BT}{E_F}\right)^2 \left(\frac{e^2}{\kappa \hbar v_F}\right)^2 I_2(q_sL) ,
\end{equation}
where $v_F$ is the Fermi velocity.
Notice that, at variance with the (side-jump) spin Hall drag resistivity, the present result is proportional to the momentum relaxation time $\tau$ and therefore it is parametrically dominant in the ``clean" limit, $\tau \to \infty$.

For the following values of the parameters: electron density
$n=2\times 10^{11}$~cm$^{-2}$, $\tau=40$~ps (which corresponds to a
mobility $\mu=10^6$ cm$^2/$Vs in a GaAs quantum well with
$m=0.067m_0$), bulk Dresselhaus splitting constant
$\gamma_c=20$~eV~\AA$^3$, $\lambda^2=5$~\AA$^2$, and $\kappa=13$, we
obtain $(\lambda k_F)^2 \approx 6\times 10^{-4}$, $\Omega_3\tau
\approx 0.9$,  and $e^2/\kappa \hbar v_F\approx 1$. Thus, taking
$I_2\sim 1$, we have $\rho_s \sim {4}
\left(\frac{k_BT}{E_F}\right)^2$~Ohm, which for $(k_BT/E_F)=0.1$ is
$0.04$ Ohm, i.e., about two times larger than the conservative
estimate for the spin Hall drag resistivity from side jump given in
Ref.~\onlinecite{badalyan-2009}. Such a value of spin drag
  resistivity can be detected by spin Faraday or Kerr rotation
  technique. In InAs based
  structures the spin-orbit coupling parameter $\lambda^2$ is about an
  an order of magnitude higher than in GaAs and $\Omega_3$ is also
  larger. This makes narrow-band semiconductor bi-layers particularly
  suitable for the observation of the spin current injection.

In conclusion, we have described a new coupling mechanism, partly Coulomb and partly spin-orbit, through which a spin current can be injected, or a spin accumulation induced,  in an electron layer by a regular electric current flowing in an adjacent layer.   The new coupling can play a role in the design of circuits in which an electric current must be converted into a spin current and viceversa.

\noindent{\textit{Acknowledgments --} M.M.G. and M.A.S. are grateful to RFBR and ``Dynasty'' Foundation---ICFPM for financial support. S.M.B. acknowledges support from EU Grant PIIF-GA-2009-235394, the DFG SFB 689, and the Belgium Science Policy (IAP). G.V. acknowledges support from NSF Grant No. DMR-0705460.


\begin{thebibliography}{90}
\bibitem{footnote} We note that an inhomogeneous Rashba interaction also arises from impurities in the doping layer,  and was previously shown~{\cite{averkiev02,sherman_rev}} to be part of the electron-impurity spin-orbit coupling.
\bibitem{note:tar} In monolayers similar arguments for electron-impurity and electron-phonon scattering were presented in Ref.~\onlinecite{tarasenko06e}.
\end{thebibliography}
 \end{document}